# NEW APPROACH TOSCALING RULES
# FOR SOLAR AND PLANETARY DYNAMOS

*Bertrand C. Barrois[1]*

Glorified dimensional analysis is used to derive scaling rules for internal and external magnetic field strengths and various time scales. Naive dimensional analysis is inconclusive because of multiple time scales, but physical arguments serve to weed out irrelevant parameters. Time scales can be derived from linearized instability analysis instead of ill-founded assumptions of Magnetic-Archimedean-Coriolis (MAC) balance. Further relationships can be derived from high-level models of coupled main field components and differential rotation. The ratios of the external dipole field to internal magnetic fields and of differential to overall rotation depend on details of the dynamo mechanism.

The dynamo mechanism by which turbulent convection sustains the magnetic fields of stars and planets has never been explained in satisfactory detail, but it is slowly yielding to numerical simulation. The goal of the present paper is to derivepredictive scaling rules for internal and external magnetic field strengths and important time scales in terms of physically relevant parameters.

Numerous authors (e.g., Christensen 2006, 2009, 2010; Davidson 2013; Oruba 2014) have addressed the scaling problem by various theoretical and numerical methods, but no consensus has emerged. There appear to be significant differences between stars and planets, as well as between fast and slow rotation regimes.

Naive dimensional analysis is inconclusive because there are no less than*four*different combinations with units of *time*:

- Rotation period of the body
- Inductive decay time for the magnetic dipole field (due to diffusion)
- Mechanical time scale derived from the luminosity (or other power source)
- Magnetic time scale derived from the magnetic energy density

We shall seek a scaling law of the general form: $T_B \sim T_\Omega^{???} \, T_Q^{???} \, T_\sigma^{???} \, f(\ldots/\ldots)$

Table 1 compares the values of these time scales in the Earth and Sun, but consistent comparisons are difficult because conditions vary with depth and because many values are uncertain.[2]Sub-surface magnetic fields and fluid flows are largely inaccessible, even though solar flares and helioseismology are revealing.

---

[1]    E-mail:  BBarrois@verizon.net

**Table 1.  Comparison of Key Parameters**

| Parameter | Formula | Earth | Sun |
|---|---|---|---|
| Outer radius | $R_{outer}$ | 3,480 km | 700,000 km |
| Inner radius | $R_{inner}$ | 1,220 km | 500,000 km |
| Depth of shell | $Z \equiv R_{outer} - R_{inner}$ | 2,260 km | 200,000 km |
| Mass of shell | $M$ | 1.15 E23 kg | 5.3 E28 kg |
| Mean density | $\overline{\rho}$ | 11.2 g/cm³ | 0.058 g/cm³ |
| Power source | $Q$ | ~ 16 E12 watts | 3.85 E26 watts |
| Polar dipole field | $B_{ext}$ | ~ 5 gauss | ~ 1 gauss |
| Differential period | $\Delta T_{\Omega}$ | ??? | 10 days |
| Rotation period | $T_{\Omega} \equiv 2\pi / \Omega$ | 1 day | ~ 30 days |
| Magnetic energy time scale | $T_B \equiv Z \sqrt{\mu \overline{\rho} / B_{ext}^2}$ | 17 years | 540 years |
| Heat transport time scale | $T_Q \equiv \sqrt[3]{MZ^2 / Q}$ | 1.05 years | 0.056 years |
| Inductive decay time scale | $T_{\sigma} \equiv \mu \sigma R_{inner}^2 / \pi^2$ | < $10^4$ years | > $10^9$ years |
| Dissipation number | $N_D \equiv \int dr \, ag / C_P$ | < 0.5 | 6.0 |

One possible clue to differences between the Earth and Sun is the ratio $T_Q / T_{\Omega}$ -- roughly 400 for the Earth, but only 0.7 for the Sun.  Other possibly significant differences involve the dissipation number and the thickness of the convective shell relative to radius.

Besides time scales, there are dimensionless parameters of possible relevance:

- The "dissipation number" for convective heat transport, which says how many times heat supplied at the base of the convective shell can be converted to kinetic energy by buoyancy and then redissipated on its way out.

- The ratio of inner to outer radii of the convective shell

- Ratios among diffusion coefficients (e.g., Prandtl number)

The lore of turbulent phenomena teaches that at extreme Reynolds numbers, kinematic viscosity (i.e., velocity diffusion) is virtually irrelevant in the inertial regime. By the same token, magnetic diffusion ($\chi_B = 1 / \mu \sigma$) ought to be irrelevant if the time

---

[2]    The magnetic susceptibility, heat capacity, thermal expansion, and transport properties of molten iron mixtures under extreme conditions are all quite uncertain and must be estimated via density functional theory.  (Anderson 1994, Alfè 2002, Pozzo 2014)

scale for inductive decay is very long. The electrical conductivity seems relevant only to boundary conditions.

Within the Sun, the electrical conductivity of plasma scales as $(Temperature)^{3/2}$, according to the Spitzer-Härm relationship (Spitzer 1954, Mitchner 1973).The diffusion time is so long that time-dependent fields cannot penetrate the non-convective inner zone. In the Earth, the conductivity of molten iron mixtures is uncertain, but it seems that the main dipole field can penetrate the solid inner core between infrequent reversals, while transient fields induced by convective eddies cannot. Recent estimates of conductivity in the outer core (Pozzo 2014) exceed 1.5 MS/m = 1.5 E6 mho/m.

*Glorified* dimensional analysisseeks to identify relevant combinations and to dismiss irrelevant parameters by appealing to physical arguments. It makes short work of three more time scales:

- Anything involving the speed of light: $R/c$
- Anything involving gravity, aside from buoyancy: $\sqrt{R/g}$
- The time for charge separation to cancel electric fields: $\varepsilon_o/\sigma$

## An Empirical Relationship

Blackett (1947) proposedanempirical relationship derived from data on the magnetically active bodies in our solar system. He found that their magnetic dipole momentsare proportional to their angular momentum, but thisrule is dimensionally unacceptable because these quantities have unlike units and there is no way to express their ratio in terms of relevant universal constants.

- Magnetic moment: $Mass^{1/2}\ Length^{5/2}\ Time^{-1}$
- Angular momentum: $Mass^1\ Length^2\ Time^{-1}$

However flawed, Blackett's curve-fit may have gotten one key point right. It is entirely possible that $B_{ext} \sim \Omega$, or equivalently, $T_B \sim T_\Omega$, other parameters being equal.

## Critique of MAC and IAC balance

A number of authors have attempted to derive scaling rules from assumptions of force or work balance, but many such assumptions lack firm physical foundation. The only rigorous principles of balance are those that derive from three related observations:

- If force terms sum to zero, and if two terms are known to dominate, then those two terms must be roughly equal and opposite. (For example, geostrophic balance assumes equality of pressure gradients and Coriolis forces, but note that such forces do no work.)

- If kinetic energy is to remain bounded, then work done on the fluid by buoyancy must over time balance work undone by magnetic and viscous forces.
- If the sum of kinetic and magnetic energy is to remain bounded, then the work done by buoyancy must over time balance the energy lost by dissipation as heat.

It is correct but uninformative to equate the work done by buoyancy to the sum of viscous and ohmic dissipation because these diffusive processes operate at vastly shorter length scales, which are not known a-priori.

Magnetic-Archimedean (MA) balance is based on the second observation. It neglects viscosity and equates work done by magnetic (Lorentz) forces to work done by buoyant (Archimedean) forces, which can in turn be related to convective heat transport. The relationship can be made rigorous by adjusting for the ratio of magnetic to total dissipation, which is *assumed* to be $o(\frac{1}{2})$, supposing rough equipartition of magnetic and kinetic energy in high-order modes and comparable diffusion coefficients:

$$\mathbf{V} \cdot [\sigma(\mathbf{E} + \mathbf{V} \times \mathbf{B}) \times \mathbf{B}] = o(1)\, V_r \frac{ag}{C_P} \tilde{H} = o(\tfrac{1}{2}) \frac{ag}{C_P} F$$

It is a serious (but common) mistake to omit the electric term and to summarize the resulting scaling rule as $\sigma V^2 B^2 \sim \frac{ag}{C_P} F$. By itself, the magnetic term opposes transverse motions and is inherently dissipative, but in most cases, the electric term offsets or cancels it via induction and/or charge separation. (For example, large-scale flows carry along small-scale magnetic field patterns without feeling an opposing force, and a metal ball can rotate freely in an applied magnetic field parallel to the axis of rotation.)

In other cases of interest, transverse fluid motions distort the magnetic field, but energy is traded elastically, as in Alfvén waves. The concept of magnetic drag is valid only when magnetic diffusion $\chi_B = 1/\mu\sigma$ has ample time to smooth the distorted field. Moreover, diffusion operates at small length scales, whereas the Kolmogorov spectrum shows that buoyant forces do most of their work at large scales.

In convective systems, genuine diffusion is overwhelmed by turbulent transport, so $\sigma B^2$ should be replaced by $B^2/\mu\chi_{eff}$. Given $\chi_{eff} \sim VZ$, the scaling rule becomes $\frac{1}{\mu} VB^2 \sim \frac{ag}{C_P} FZ$, which is equally consistent with $V \sim B \sim F^{1/3}$ and $V \sim F/\Omega^2, B \sim \Omega$.

Archimedean-Coriolis (AC) balance asserts that buoyant forces scale with Coriolis forces, but the physical premise of AC balance is suspect because pressure gradients are omitted and Coriolis forces do no work.

$$\frac{ag}{C_P}\tilde{H} \sim \Omega V \quad \Rightarrow \quad \frac{ag}{C_P}F \sim \Omega V^2$$

Inertial-Archimedean-Coriolis (IAC) balance asserts that inertial forces scale with buoyant forces, even though inertial forces do no work overall, and also with Coriolis forces. The derivation in (Aubert & Brito 2001) can be summarized as follows:

$$(\mathbf{V}\cdot\nabla)\mathbf{V} \sim \frac{ag}{C_P}\tilde{H} \sim \Omega V$$

$$V^3/l^* \sim \frac{ag}{C_P}F \sim \Omega V^2 \, l^*/Z$$

If $\Omega = 0$, then $l^* \sim \min(Z, Z/N_D)$ and $V \sim F^{1/3}$; but given rapid rotation, $V \sim F^{2/5}\Omega^{-1/5}$. The latter prediction combines with MA balance to give $B \sim F^{3/10}\,\Omega^{1/10}$, per (Jones 2010).

**Lessons from Instability Analysis**

Convective instability results from a positive-feedback loop, in which vertical (radial) flows carry warmer fluid into cooler environments and buoyancy drives said flows. Disregarding horizontal (tangential) components and spatial variation of the flow, in the cavalier spirit of dimensional analysis, we may derive the *limiting* exponential growth rate by eigenvalue analysis of a linearized system:

$$\frac{d}{dt}\begin{bmatrix}\tilde{H}\\ V_r\end{bmatrix} = \begin{bmatrix}0 & \bar{H}'_>\\ \frac{ag}{C_P} & 0\end{bmatrix}\begin{bmatrix}\tilde{H}\\ V_r\end{bmatrix}$$

The limiting eigenvalues are $\Gamma_0 = \pm\sqrt{\frac{ag}{C_P}\bar{H}'_>}$, where $\bar{H}'_> \equiv \left|\frac{d}{dz}\bar{H}\right| - g > 0$ denotes the super-adiabatic gradient of enthalpy per unit mass. This relationship is not very useful if the gradient is unknown, but its scaling rule is obvious. Let $F$ denote the heat transported per unit area through a flat convective layer, and $Z$ the distance between the boundaries. Dimensional analysis demands that $\Gamma_0 \sim \sqrt[3]{N_D F/\rho Z^3}$ for $N_D < 1$.

The case where $N_D > 1$ is relevant to the Sun, but data are scarce because in most experiments on heat transport, $N_D$ is held constant, while $F$ and $\Delta T$ vary. Since it is not possible to convert more than 100% of the heat supplied from below to kinetic energy, we shall use $\Gamma_0 \sim \sqrt[3]{F/\rho Z^3}$ when $N_D > 1$. We may identify $T_Q = 1/\Gamma_0$.

Both Coriolis forces and the main magnetic field act as restraints to radial convection. Stronger temperature gradients are needed to create instability, and in most cases, exponential growth rates are reduced.

Supposing that Coriolis forces convert $V_X \rightarrow V_Y \rightarrow -V_X$ in a flat layer, we could work them into the eigenvalue problem as follows:

$$\frac{d}{dt} \begin{bmatrix} \tilde{H} \\ V_X \\ V_Y \end{bmatrix} = \begin{bmatrix} 0 & \bar{H}'_> & 0 \\ \frac{ag}{C_P} & 0 & +\Omega_Z \\ 0 & -\Omega_Z & 0 \end{bmatrix} \begin{bmatrix} \tilde{H} \\ V_X \\ V_Y \end{bmatrix}$$

When $\exp(i\mathbf{k} \cdot \mathbf{x})$ spatial variation is assumed, the relevant eigenvalues are found to be $\Gamma(\mathbf{k}) = \pm \sqrt{\left(1 - (\hat{\mathbf{g}} \cdot \hat{\mathbf{k}})^2\right) \frac{ag}{C_P} \bar{H}'_> - (\boldsymbol{\Omega} \cdot \hat{\mathbf{k}})^2}$, where hats indicate unit directional vectors. Rapid rotation is seen to suppress the convective instability when $\mathbf{k}$ is parallel to $\boldsymbol{\Omega}$, in accord with the Taylor-Proudman theorem. In spherical geometry, rotation preferentially restrains low-M convection modes that vary rapidly with respect to latitude.

There is no satisfactory theory to relate the exponential growth rates inferred from linearized instability analysis to the distribution of energy in convectively driven systems (Barrois 2016a,b). Rigorous dimensional scaling rules *à la* Kolmogorov could only be stated if all growth rates were rescaled by the same factor.

We cannot conclude that $\frac{ag}{C_P} \bar{H}'_> > \Omega^2$ because a subset of convective modes with $\mathbf{k}$ roughly perpendicular to $\boldsymbol{\Omega}$ remains unstable. The driven subset comprises a fraction of all modes: $f \sim \sqrt{\frac{ag}{C_P} \bar{H}'_> / \Omega^2} = \Gamma_{max} / \Omega$. Since the heat transport $F \sim f \Gamma_{max} Z \rho \bar{H}'_>$ is a fixed quantity, inclusion of the factor $f < 1$ enhances the gradient and the maximum growth rate: $\Gamma_{max} = \sqrt{\frac{ag}{C_P} \bar{H}'_>} \sim \Gamma_0^{3/4} \Omega^{1/4}$ but $f \Gamma_{max} \sim \Gamma_0^{3/2} \Omega^{-1/2}$ .

By the same token, slowly-varying magnetic fields oppose high-$\mathbf{k}$ transverse fluid motions, with $\frac{1}{\sqrt{\mu\rho}} (\mathbf{B} \cdot \mathbf{k})$ in place of $(\boldsymbol{\Omega} \cdot \hat{\mathbf{k}})$. Energy is traded back and forth between the fluid and distorted fields in elastic fashion, setting up Alfvén waves. In spherical geometry, the main toroidal field preferentially restrains high-M convection modes that vary rapidly with longitude, and which are essential to dynamo mechanisms, per Cowling's theorem. The key modes are presumed to have azimuthal wavenumbers of order $\frac{\pi}{Z}$ .

If we unify the treatment of these two mechanisms and define $\Psi \equiv \frac{\pi}{Z\sqrt{\mu\rho}} B_{int}$ , then $f = \Gamma_{max}^2 / \Omega \Psi$ and $\Gamma_{max} = \sqrt{\frac{ag}{C_P} \bar{H}'_>} \sim \Gamma_0^{3/5} \Omega^{1/5} \Psi^{1/5}$ but $f \Gamma_{max} \sim \Gamma_0^{9/5} \Omega^{-2/5} \Psi^{-2/5}$ .

By itself, this relationship does not predict $V$ or $B$, but it can be combined with $V \sim Z f \Gamma_{max}$ and the quasi-rigorous prediction of MA balance:

$$\tfrac{1}{\mu} V B^2 \sim \frac{ag}{C_P} F Z$$

$$B_{rms} \sim \Gamma_0^{-3/4}\ \Omega^{1/4}$$

$$V_{rms} \sim \Gamma_0^{-3/2}\ \Omega^{-1/2}$$

**Non-uniform gradients**

The gradient is actually non-uniform, so the eigenvalue problem involves a differential equation with boundary condition $V(0) = V(Z) = 0$.

$$\Gamma^2(k_T^2 - \partial_Z^2)V(z) = k_T^2 \bar{H}_>'(z)\,V(z)$$

One plausible heuristic from the theory of heat transport defines a *local* turbulent transport coefficient $\chi_{eff}(z) \sim z^2\,\Gamma$, where $z$ relates to distance from the nearer boundary. The overall super-adiabatic temperature difference would diverge, were it not for diffusive heat conduction through the boundary layers:

$$\Delta T = \frac{2F}{\rho C_P} \int_0^{Z/2} \frac{dz}{\chi_{eff}(z) + \chi_H} \approx \frac{\pi F}{\rho C_P \sqrt{\Gamma \chi_H}} \sim F^{5/6}$$

Actual temperature profiles have been explored experimentally, but the relationship between $F$ and $\Delta T$ is shaky. In terms of the Rayleigh number $Ra \equiv ag\,\Delta T\,Z^3 / \chi_H \chi_V$, and Nusselt number $Nu \equiv FZ / \chi_H C_P \Delta T$, local $\chi_{eff}(z) \sim z^2 \Gamma$ suggests $Nu \sim Ra^{1/5}$, whereas global $\chi_{eff} \sim Z^2 \Gamma$ suggests $Nu \sim Ra^{1/2}$. Experiments cited by (Faber 1995) have found $Nu \sim Ra^{0.33}$ for $Ra < 10^8$, but $Nu \sim Ra^{0.28}$ for $Ra > 10^8$.

If the profile is very steep within boundary layers, but flat elsewhere, then the eigenfunction $V(z)$ can also be relatively flat, except in the boundary layers.

**Stratified density**

If the density varies with depth, then the equation is altered:

$$\Gamma^2(k_T^2 - \partial_z \rho^{-1} \partial_z \rho^{+1})\,V(z) = k_T^2 \bar{H}_>'(z)\,V(z)$$

If the density varies very slowly, then $V(z)$ is rescaled by a factor of $\rho^{-1/2}$, and the distribution of kinetic energy is essentially unaffected. Near the surface of a star, $\rho^{2/3} \sim T \sim \Phi(R) - \Phi(r)$, where $\Phi$ denotes the gravitational potential. The scale height is ill-defined.

**Lessons from a high-level dynamo model**

An earlier paper by the author (Barrois 2015) proposed a model of the solar dynamo in which three basic processes cooperate to cause oscillations and drive differential rotation. The familiar *omega* effect bends poloidal field lines $(B_S)$ into toroidal hairpins $(B_T)$, a multi-step regeneration process termed the *zeta* effect regenerates the dipole field, and their cross-term drives differential rotation $(\Omega_D)$. The omega effect depends ondifferential rotation, whereas the *zeta* effect depends on overall rotation $(\Omega_C)$ via Coriolis forces on eddies. Linear damping terms and additive noise drivers $N_{...}(t)$ are included to represent the influence of higher-multipole modes omitted from the three-variable model.

$$\tfrac{d}{dt} B_S = +\zeta \, \Omega_C \, B_T - g_S B_S + N_S(t) \quad (a)$$
$$\tfrac{d}{dt} B_T = -\omega \, \Omega_D \, B_S - g_T B_T + N_T(t) \quad (b)$$
$$\tfrac{d}{dt} \Omega_D = +\omega \, B_S \, B_T - g_D \Omega_D + N_D(t) \quad (c)$$

If we drivers and eliminate $\Omega_D / B_T$ from equations (b) and (c), we conclude that $\omega^2 B_S^2 = g_D g_T$ in a quasi-constant state. Given damping rates $g \sim f\Gamma_{\max}$, we are led to the scaling rule $B_S \sim f\Gamma_{\max}$. Similar manipulation of equations (a) and (b) leads to $\zeta \omega \Omega_C \Omega_D = -g_S g_T$, which may be correct for the Earth, but oscillatory systems such as the Sun violate the premise of a quasi-constant state.

The bird's-eye model predicts that $\Omega_D$ will level off provided that $\langle B_S B_T \rangle = 0$. A long, wooly, and ultimately inconclusive analysis relates $\langle \mathbf{BB} \rangle$ to $\langle \mathbf{NN} \rangle$, which must in practice be extracted from detailed numerical simulations. The response to a noise process with time dependence $\exp(iwt)$ may be calculated as follows:

$$\mathbf{B} = \begin{bmatrix} B_S \\ B_T \end{bmatrix} = \begin{bmatrix} iw + g_S & -\zeta \, \Omega_C \\ +\omega \Omega_D & iw + g_T \end{bmatrix}^{-1} \begin{bmatrix} N_S \\ N_T \end{bmatrix} = \mathbf{M} \, \mathbf{N}$$

$$\langle \mathbf{BB} \rangle = \mathbf{M} \langle \mathbf{NN} \rangle \mathbf{M}^\dagger$$

Hypothetically, if $\langle \mathbf{NN} \rangle$ were simply a multiple of the identity matrix, then one could conclude that $\omega \Omega_D g_S = \zeta \, \Omega_C g_T$, but in fact, $\langle N_S N_T \rangle \neq 0$ in rotating systems.

Anyconclusion to the effect that $\Omega_D \sim \Omega_C^{\pm 1}$ seems surprising in terms of tensor character. Overall rotation has dipole (L=1) character, whereas differential rotation has octopole (L=3) character.

There is a further mechanism by which a convectively driven flow mode can pump a weakly damped magnetic field mode, such asa main field component. An analysis of hyperbolic sloshing in (Barrois 2016a) suggests that the amplitude of the pumped magnetic field component should scale with the growth rate of the unstable flow mode.

**Comparison with numerical results**

The foregoing arguments suggest the following rules for the *slow*-rotation regime, relevant to the Sun:

- Lorentz number:     $B_{rms} \sim \Gamma_0 \sim Q^{1/3}$
- Rossby number:     $V_{rms} \sim \Gamma_0 \sim Q^{1/3}$

But for the *rapid*-rotation regime relevant to the Earth, the theoretical arguments bog down in unvalidated heuristics, which suggest:

- Lorentz number:     $B_{rms} \sim Q^{1/4} \; \Omega^{1/4}$
- Rossby number:     $V_{rms} \sim Q^{1/2} \; \Omega^{-1/2}$

The externally observable poloidal dipole field $B_{ext}$ may scale like $V_{rms}$, and differential rotation may be inversely related to overall rotation.

Christensen & Aubert (2006) extracted scaling rules from numerical simulations of Earth-like dynamos in the *rapid*-rotation regime. Little differential rotation was observed, and the poloidal dipole and toroidal quadrupole components were found comparable.

- Lorentz number:     $B_{rms} \sim Q^{0.34} \; \Omega^{-0.02}$
- Rossby number:     $V_{rms} \sim Q^{0.41} \; \Omega^{-0.23}$

This surprising scaling rule for the Rossby number agreed with experimental results of Aubert & Brito (2001) and their theoretical rationalization in terms of IAC balance.